# The signature of robot action success in EEG signals of a human observer: Decoding and visualization using deep convolutional neural networks

Joos Behncke[1,2], Robin T. Schirrmeister[1,2], Wolfram Burgard[1], Tonio Ball[2]

[1] Department of Computer Science, Albert-Ludwigs-University Freiburg, Germany

[2] Translational Neurotechnology Lab, University Medical Center Freiburg, Germany

joos.behncke@uniklinik-freiburg.de

*Abstract*—The importance of robotic assistive devices grows in our work and everyday life. Cooperative scenarios involving both robots and humans require safe human-robot interaction. One important aspect here is the management of robot errors, including fast and accurate online robot-error detection and correction. Analysis of brain signals from a human interacting with a robot may help identifying robot errors, but accuracies of such analyses have still substantial space for improvement. In this paper we evaluate whether a novel framework based on deep convolutional neural networks (deep ConvNets) could improve the accuracy of decoding robot errors from the EEG of a human observer, both during an object grasping and a pouring task. We show that deep ConvNets reached significantly higher accuracies than both regularized Linear Discriminant Analysis (rLDA) and filter bank common spatial patterns (FB-CSP) combined with rLDA, both widely used EEG classifiers. Deep ConvNets reached mean accuracies of 75% ± 9 %, rLDA 65% ± 10% and FB-CSP + rLDA 63% ± 6% for decoding of erroneous vs. correct trials. Visualization of the time-domain EEG features learned by the ConvNets to decode errors revealed spatiotemporal patterns that reflected differences between the two experimental paradigms. Across subjects, ConvNet decoding accuracies were significantly correlated with those obtained with rLDA, but not CSP, indicating that in the present context ConvNets behaved more "rLDA-like" (but consistently better), while in a previous decoding study with another task but the same ConvNet architecture, it was found to behave more "CSP-like". Our findings thus provide further support for the assumption that deep ConvNets are a versatile addition to the existing toolbox of EEG decoding techniques, and we discuss steps how ConvNet EEG decoding performance could be further optimized.

## I. Introduction

Recently, there is a great interest in assistive robotic solutions in healthcare and in non-medical applications. Intelligent robotic systems have a large potential to facilitate working processes and to endow persons with limited motor abilities with more autonomy. However, collaborative scenarios, especially when robots share the same workspace with a human user, require a safe management of "robot errors", e.g., when robot behavior disagrees with the user's intentions.

While it would be optimal to prevent such robot errors entirely, this is unlikely to become feasible soon. Thus, detection of robot errors and correction of their consequences remains a relevant problem.

Brain signals may be helpful (as a source of information in healthy robot users as well as in neurological patients with severely limited communication abilities) to this purpose. In the last years, several studies investigated decoding of robot errors from brain signals of a human observer [1,2,5]. Error-related potentials recorded with EEG have been used, e.g., to teach neuroprosthetics suitable behaviors in scenarios of varying complexity [1] or to investigate their role for robot control during an object selection task [2]. Accuracies however leave space for improvements, which would be desirable to optimize the practical usefulness of error-related brain signals.

We addressed this problem by applying deep learning to a naturalistic decoding task where participants observed a robot performing different assistive actions either successfully or failing to do so. In EEG research, architectures including deep Convolutional Neural Networks (ConvNets) have recently been used to explore their applicability in brain-signal decoding [3,4], but not yet to robot-error decoding from EEG.

## II. System and Experimental Design

To investigate the detection of error related EEG activity, we designed two experiments where participants observed robots performing naturalistic actions either in a correct or an erroneous manner. In both experiments, visual stimuli were short video clips, repeatedly presented in a randomized order. In Fig. 1 the experimental paradigms are schematically shown, Fig. 2 depicts the trials structure. [5] describes the setup in detail.

### A. Experiment 1: KUKA Pouring Observation (KPO)

In this experiment, video stimuli showed a robotic arm (LBR IIWA, KUKA, Augsburg, Germany) pouring liquid from a non-transparent bottle either successfully into a glass, or spilling the liquid. The two conditions were held equal as far as possible (same camera and glass position, same robot movement). There was either a greater or a smaller amount of liquid inside the bottle, which was not visible for the subject. This alone determined whether the outcome of the pouring video stimulus was successful or not (the greater amount of liquid let to spilling as the liquid would overshoot the target vessel).

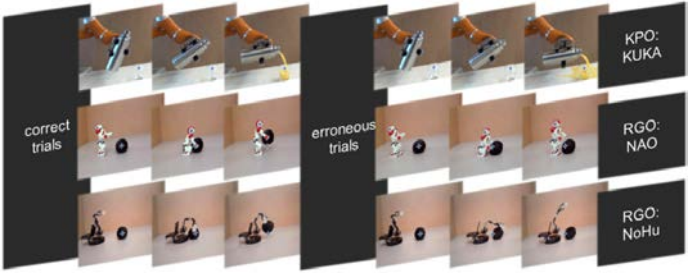

Figure 1. Experimental procedure. Schematic of the trial structure and video stimuli used in the experiments. For both robot types, there was a correct and incorrect condition. In the experiment 1 (KPO), a robotic arm performed a pouring task, either hitting or missing the vessel. In experiment 2 (RGO), either a humanoid robots (NAO) or a non-humanoid robot (NoHu) performed a grasping task, either managing or failing to lift a ball from the ground.

### B. Robot-Grasping Observation (RGO)

Here, a humanoid (NAO) and a non-humanoid robot (referred to as NoHu) were programmed to approach, grab, and lift a ball from the ground, again either succeeding or failing (letting the ball drop from the grasper). Again, the videos were invariant concerning starting position of the robot, initial position of the ball, and visual properties of the surrounding. The videos were cut to maintain the same time duration and time structure across conditions and robot types. To generate clips showing the robots approaching from left and right, the existing ones were vertically flipped. The RGO paradigm thus enables not only the investigation of error-related signals, but also of differential responses elicited by the two types of robots, which was however not the aim here.

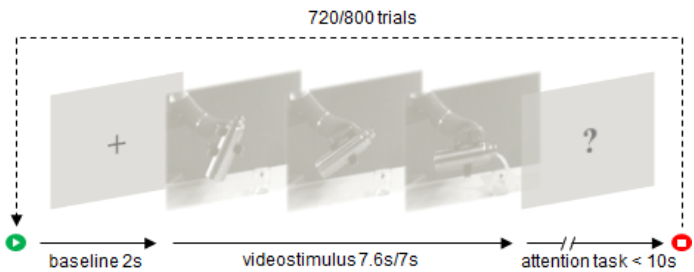

Figure 2. Timing structure of the experimental trials, comprising video stimuli of 7.6-s (KPO) and 7-s (RGO) durations and an attention control task. Each participant completed 720 (KPO) and 800 (RGO) trials, respectively.

### C. Participants and EEG Acquisition

All participants were healthy and providing their informed consent for the study, which was approved to by the local ethics committee. Number of participants (22-31 years old) was 5 (5 male) for KPO and 12 (6 female) for RGO. Participants were comfortably seated inside a dimly lit electro-magnetically shielded EEG recording chamber. EEG was acquired with an EEG cap with 128 gel-filled electrodes positioned according to the 'five percent' electrode layout. All participants were instructed to fix their gaze onto a fixation cross (see Fig. 2) before each video was started and were asked to keep the fixation during the stimulus. The subsequent attention task permitted the participants to move slightly, blink or swallow, while maintaining their attention during the experiment. Additionally, we implemented an extra condition in the RGO task, where in 10% of the trials the participants were instructed to press a button if and exactly when they perceived a robot error. This allowed an estimation of the approximate time point of error perception. The experiment was conducted in sessions of 30 trials (KPO) and 40 trials (RGO), respectively. Altogether at least 720 and 800 trials per participant were recorded in experiment KPO and RGO, respectively.

### III. SIGNAL PRE-PROCESSING AND CLASSIFIER DESIGN

The recorded EEG datasets were re-referenced to a common average (CAR) and resampled to 250 Hz. To compute exponential moving means and variances for the ConvNets, an electrode-wise exponential moving standardization with a decay factor of 0.999 was applied [3], while the rLDA implementation reached higher accuracies without the standardization. Based on predefined decoding intervals, the data was cut into trials according to the stimulus onset. Pre-processing and implementation of the FB-CSP algorithm, following [6], is discussed in detail in [5]. Data analyses employing rLDA and deep ConvNets are based on python implementations, with the deep ConvNets obtained from *braindecode*, an open-source deep learning toolbox for raw time-domain EEG. The underlying architecture is described in [3], training and classification was only done within each participant. The number of layers is the same as in the original paper whereas a stride of 2 was used to ensure a smaller receptive field. The architecture of the rLDA classifier complies with the theory of [7] and is leant on the realization of [8], for the shrinkage regularization we used the "LedoitWolf" estimator [9].

### IV. STATISTICS

Significance for individual decoding results was estimated using a permutation test [10, 11]. To create the null distribution a randomization process of n=1000000 guesses was applied. For each guess the genuine classification labels were randomly assigned to the trials and the number of correct classifications was counted. The yielding probability distribution was utilized as a base for the estimation of the significance of the achieved decoding accuracies. Mean differences of accuracies between decoding methods were evaluated by Wilcoxon signed-rank tests [12]. Significance of correlation coefficients was evaluated by randomizing the order of one of the input vectors of the correlation. The number of guesses that resulted in higher coefficients than the true correlation coefficient was compared to the total number of guesses.

TABLE I. COMPARISON OF DECODING PERFORMANCES

| paradigm | interval | mean accuracy ± standard deviation | | |
|---|---|---|---|---|
| | | *ConvNet* | *rLDA* | *FB-CSP* |
| KPO error | 2.5-5s | (78.2 ± 8.4) % | (67.5 ± 8.5) % | (60.1 ± 3.7) % |
| KPO error | 3.3-7.5s | (71.9 ± 7.6) % | (63.0 ± 9.3) % | (66.5 ± 5.7) % |
| RGO error | 4.8-6.3s | (59.6 ± 6.4) % | (58.1 ± 6.6) % | (52.4 ± 2.8) % |
| RGO error | 4-7s | (64.6 ± 6.1) % | (58.5 ± 8.2) % | (53.1 ± 2.5) % |

## V. COMPARISON OF DECODING PERFORMANCE

We implemented three different decoding algorithms (ConvNets, rLDA, FB-CSP+rLDA) and compared the outcomes (Tab. 1). We compared results for three different time intervals defined in the EEG data from both experiments as also indicated in Tab. 1. For both experiments and all time intervals, ConvNets yielded the highest decoding accuracies.

Fig. 3 shows the pairwise comparison between decoding accuracy obtained in the individual subjects for all three methods for error decoding. The corresponding mean accuracies are listed in Table 1. The decoding intervals 3.3-7.5s (KPO) and 4-7s (RGO) were selected according to [5]. Additionally, for KPO we analyzed the data between 2.5-5s, as this seemed as an intuitive interval in which the error became obvious. As described before, in the RGO paradigm we integrated an extra condition to serve as an error perception feedback. We evaluated the responses and selected the 5-fractile range of the time values of all button press events, resulting in the decoding interval 4.8-6.3s.

Fig. 3 shows that in KPO the ConvNet decoding accuracies significantly exceeded those of the other two decoding methods for each single participant, and on the group level was significantly better compared to both rLDA and FB-CSP. There was no significant difference between the two latter methods in KPO on the group level, however, there were significant differences between rLDA and CSP on the individual level which were almost always in favor of rLDA (Fig. 3 left column, bottom panel).

TABLE II. LINEAR CORRELATION OF DECODING PERFORMANCES

| Paradigm | Linear correlation coefficient (p-value) | | |
| --- | --- | --- | --- |
| | *ConvNet/rLDA* | *ConvNet/FB-CSP* | *rLDA/FB-CSP* |
| KPO error | 0.913 ($<$ 0.001) | 0.292 (0.213) | 0.375 (0.152) |
| RGO error | 0.512 (0.004) | 0.277 (0.095) | 0.162 (0.223) |

In RGO, comparing ConvNets to rLDA, significant differences were also nearly in all cases in favor of the ConvNets. In contrast to KPO, in part of the subjects there was no significant performance difference detectable, and there was also no significant difference on the group level. Compared to FB-CSP, ConvNets were however again significantly better in nearly all individual subjects and also on the group level. In RGO (but not in KPO), rLDA significantly outperformed FB-CSP.

To quantize the relationship between the methods, we calculated the linear correlation between decoding accuracies over subjects, pairwise for the different methods. Fig. 4 shows the correlation for the comparison of the ConvNet and the rLDA performances for KPO and RGO error decoding. Particularly for the error decoding in KPO, there was a highly significant linear correlation. There was no significant correlation with FB-CSP performance. Results are summarized in Table 1.

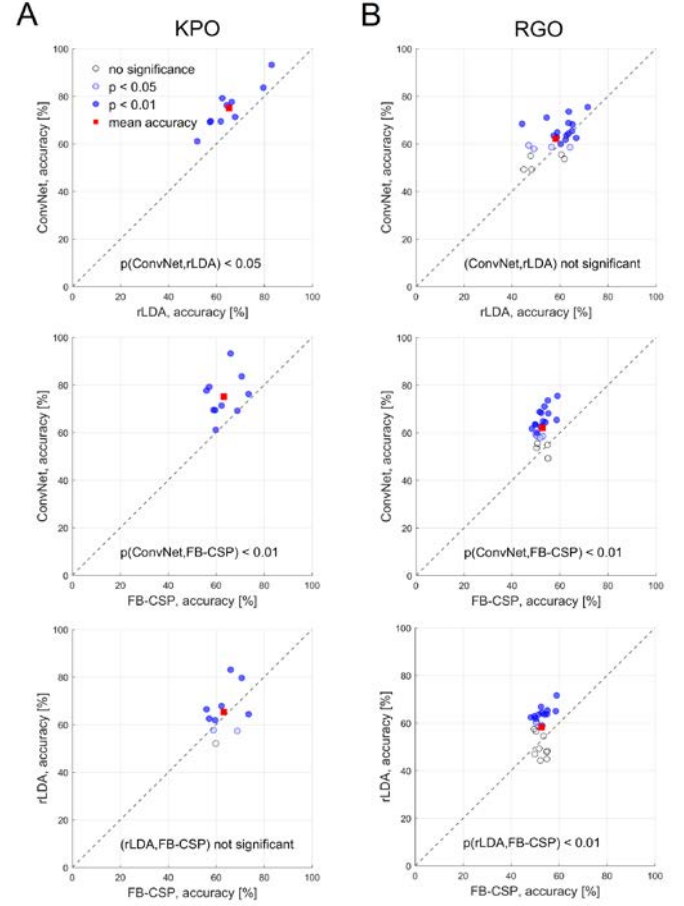

Figure 3. Pairwise comparison of decoding performance of the three different methods investigated.

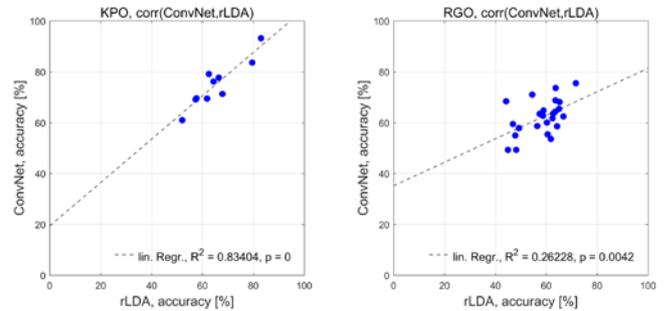

Figure 4. Correlation of ConvNet and rLDA results for both paradigms.

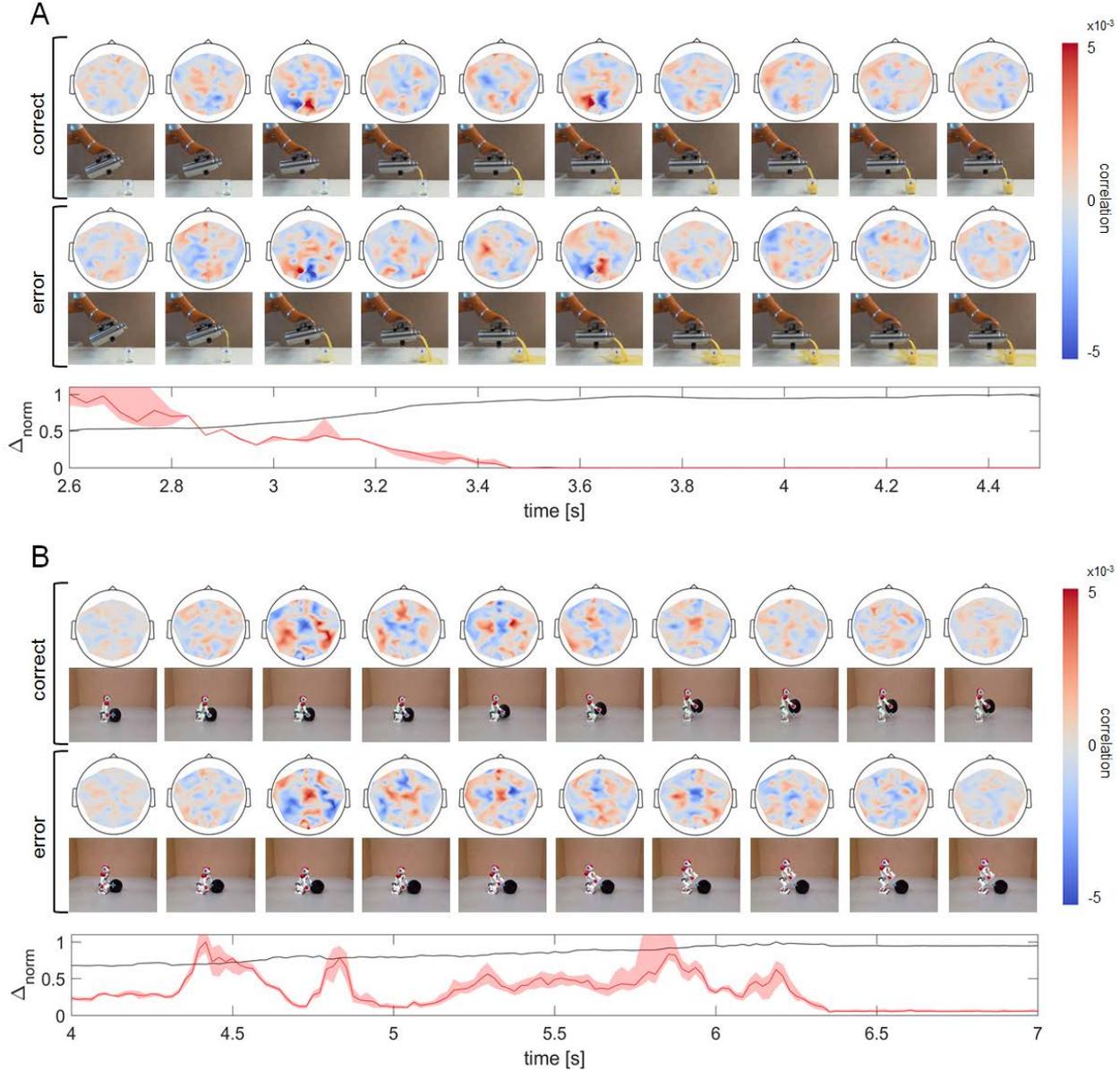

Figure 5. **A** Time-resolved voltage feature input-perturbation network-prediction correlation maps for error decoding, averaged over 30 iterations and all KPO participants (top). Time-resolved normalized L1 distance $\Delta_{norm}$ between (1) video frames for both conditions (bottom, black) and of (2) sequential pairs of video frames for both conditions (bottom, red). **B** Visualizations for RGO error decoding, all conventions as in A.

## VI. VISUALIZATION

Firstly, we used the correlation of changes in ConvNet predictions with perturbation changes in input spectral amplitudes to obtain information about what the deep ConvNets learned from the data [3]. Training trials were transformed into frequency domain using the Fourier transformation and randomly perturbed by adding Gaussian noise, while keeping the phases steady. Both, the unperturbed and the inverse-Fourier transformed signals were used to feed the deep ConvNet. We obtained the output of the ConvNet before the softmax activation for 30 iterations and as described in [3] the difference of the perturbed and original ConvNet predictions were correlated with the perturbation itself. Secondly, we also applied perturbations to the time domain voltage signal. Again, the differences of the two signals were correlated with the ConvNet output changes.

Here, we focus on the visualization results in the time domain, as rLDA trained on the time-domain EEG signals outperformed CSP (the latter designed to exploit band-specific spectral power differences, Fig. 4) and in the RGO experiment CSP decoding was at chance level. This suggests that in the present decoding problems, band-specific spectral power differences did not play the dominant role as a source of decodable information. Accordingly, frequency-resolved ConvNet visualizations (not shown) were rather noisy-looking.

In Fig. 5A, the averaged time-resolved input-perturbation network-prediction correlation maps for voltage features of the two decoding classes in the error decoding of the KPO paradigm are shown. Video frames shown below the maps were selected according to the specific point in time of each map.

The patterns of the correct and error classes showed two times windows with high correlation, first around 3.1s and then again around 3.7s (time relative to the onset of the video stimuli). In both instants the network appears to learn a similar occipitally-pronounced EEG pattern. The comparison of maps from both conditions expectedly shows opposite patterns. The occipital predominance of correlation effects in these time windows would suggests that the subjects' brains differentially processed visual aspects distinguishing correct and incorrect robot action as presented in the stimuli.

As a first step to investigate which visual features carried the error-specific information, we calculated the L1 distance between temporally corresponding frames in both conditions, as well as between the frame-wise change (black curves in Fig 5A and B). At least with these simple features, there was no obvious relation between the time course of changes in them and the time points where the EEG was most informative for ConvNet error decoding.

Analogous visualizations for error decoding in the RGO experiment showed spatially more widespread effects (Fig. 5B). Temporally, however, these effects had a remarkable sharp onset approx. 4.75 sec into the video stimuli, around the time when success vs. failure became first evident, but long before the obvious consequences of error vs. success became visible (ball being lifted from the ground or not). Again, there was no obvious time relation to the two low-level measures of image similarity (Fig. 5B, bottom panel).

## VII. Conclusions & Outlook.

The findings of the present study can be seen from two sides: first, decoding the success of robot action from brain signals is a problem with potential practical relevance and, hence, has been investigated in a number of previous studies [1,2,5]. More specifically, improving the decoding accuracy in this context is a topic with practical relevance, particularly under complex, real-life-like conditions. Thus we designed video stimuli to mimic such conditions. Second, ConvNets are still relatively new in EEG decoding, and the findings from the present decoding problems also contribute some new facts to the growing methodological literature on this topic.

Our results show that, compared with 2 other widely-used classifier, our deep ConvNets performed consistently better. Notably, we used them "Out of the box" as previously described and evaluated in [5], to avoid overfitting to our problems. The same ConvNet as applied here yielded mean accuracies of 93% for classification of 4 different movements in [3] and 85% in discriminating normal and pathological EEG in [13], and was in all cases at least as good or significantly better than the baseline comparison methods.

In the present study we reached mean accuracies of 75% ± 9 % (KPO) and 62% ± 7% (RGO) for error decoding. In a previous study, using a combination of rLDA and reinforcement learning, decoding of actions that subjects evaluated as either erroneous or correct [1] resulted in a mean EEG decoding accuracy of 75%. For one of the paradigms (KPO) we reached a similar mean accuracy here. In some of our subjects, accuracies were above 90%, but overall still better accuracies are needed. Among other recent advances in the field of deep learning research, automatic hyperparameter optimization and architecture search, including recurrent and residual network architectures, data augmentation, using 3-D convolutions, or increasing the amount of training data all have the potential to further increase ConvNet performance.

ConvNets were systematically better but in their accuracies over subjects linearly correlated with those of the rLDA, but not of FB-CSP (Fig. 4). So in the present examples the ConvNets behaved "rLDA-like". Interestingly, in a previous study where the same ConvNet architecture and training strategy as here was used for movement decoding from EEG, accuracies over subjects were highly correlated with those of FB-CSP [3], and it was shown that ConvNets indeed used frequency-specific spectral power changes (rLDA was not evaluated there). This points to the possibility that ConvNets might become more "CSP-like" or more "rLDA-like" (or even more similar to other decoding methods) depending on what features are informative in the EEG signal.

The results as discussed so far indicated an important role of time-domain EEG signal changes for the decodability of errors in our tasks, thus for their visualization we adapted the perturbation-based technique as described in [3] for spectral changes to time-domain voltage features. Resulting maps confirmed that the ConvNets learned to use time-domain EEG responses to distinguish between classes. Maps also indicated that specific time windows and scalp regions were informative, with different patterns in the two tasks (Fig. 5). Particularly for errors in the pouring task (KPO), perturbation maps pointed to the occipital/visual areas as important sources of information learned by the ConvNets. This kind of decoding could be practically helpful in situations where robot errors would be visually distinct, such as in our example of liquid spilling to a table. Further it would be interesting to investigate in how far the decodability of such differential visual input depends on its subjective interpretation as an error.

Maps visualizing which EEG signals ConvNets learned to decode errors in the grasping task (RGO) showed a spatially more widespread pattern, but also with a relatively sharp onset around the time when failure and success became first evident from the stimuli (Fig. 5B). Speculatively, observation of the reaching-grasping task might activate the human mirror neuron system (MNS) [14-17]. The human MNS involves widespread frontal and parietal regions as involved in the maps in Fig. 5B. The engagement of the MNS might be modulated by the degree of humanoid appearance of the robot. Thus as a next step, we would analyze differences related to the two robot types (more and less humanoid) used in our reaching-grasping experiment.

ACKNOWLEDGMENTS

This research was supported by the German Research Foundation (DFG, grant number EXC1086) and grant BMI-Bot by the Baden-Wuerttemberg Stiftung.

REFERENCES

[1] I. Iturrate, R. Chavarriaga, l. Montesano, J. Minguez, and J. del R. Millan, "Teaching brain-machine interfaces as an alternative paradigm to neuroprosthetics control." Scientific reports 5 (2015).

[2] A.F. Salazar-Gomez, J. DelPreto, S. Gil, F.H. Guenther., and D. Rus, "Correcting robot mistakes in real time using eeg signals." ICRA 2017. IEEE.

[3] R.T. Schirrmeister, J.T. Springenberg, L.D.J. Fiederer, M. Glasstetter, K. Eggensperger, M. Tangermann, F. Hutter, W. Burgard, and T. Ball, "Deep learning with convolutional neural networks for EEG decoding and visualization." Human brain mapping (2017).

[4] Z. Tang, C. Li, and S. Sun, "Single-trial EEG classification of motor imagery using deep convolutional neural networks." Optik - International Journal for Light and Electron Optics (2017), 130:11-18.

[5] D. Welke*, J. Behncke*, M. Hader, R. T. Schirrmeister, A. Schönau, B. Eßmann, O. Müller, W. Burgard, and T. Ball, *equally contributing, "Brain responses during robot-error observation." Kognitive Systeme (2017), in press.

[6] K. K. Ang, Z. Y. Chin, C. Wang, C. Guan, and H. Zhang, "Filter bank common spatial pattern algorithm on BCI competition IV datasets 2a and 2b." Frontiers in neuroscience 6 (2012).

[7] J. H. Friedman, "Regularized discriminant analysis." Journal of the American statistical association (1989) 84, 165.

[8] B. Blankertz, M. Tangermann, C. Vidaurre, S. Fazli, C. Sannelli, S. Haufe, C. Maeder, L. Ramsey, I. Sturm, G. Curio, and K. R. Müller, "The Berlin brain–computer interface: non-medical uses of BCI technology." Frontiers in neuroscience, 4 (2010)

[9] O. Ledoit and M. Wolf, "A well-conditioned estimator for large-dimensional covariance matrices." Journal of multivariate analysis 88.2 (2004): 365-411.

[10] R. A. Fischer, "The coefficient of racial likeness and the future of craniometry." Journal of the Royal Anthropological Institute (1936), 66, 57-63

[11] E. J. G. Pitman, "Significance tests which may be applied to samples from any populations." Supplement to the Journal of the Royal Statistical Society (1937), 4, 119-130.

[12] F. Wilcoxon, "Individual Comparisons by Ranking Methods." Biometrics Bulletin (1945), 1(6):80-83.

[13] R. T. Schirrmeister, L. Gemein, K. Eggensperger, F. Hutter, and T. Ball, "Deep learning with convolutional neural networks for decoding and visualization of EEG pathology." unpublished.

[14] R. Hari, N. Forss, S. Avikainen, E. Kirveskari, S. Salenius, and G. Rizzolatti, "Activation of human primary motor cortex during action observation: a neuromagnetic study." Proceedings of the National Academy of Sciences (1998), 95(25), 15061-15065.

[15] G. Rizzolatti, L. Fadiga, V. Gallese, and L. Fogassi, "Premotor cortex and the recognition of motor actions." Cognitive brain research (1996), 3(2), 131-141.

[16] G. Rizzolatti, and L. Craighero, "The mirror-neuron system." Annual Review of Neuroscience (2004), 27, 169-192.

[17] R. Mukamel, A. D. Ekstrom, J. Kaplan, M. Iacoboni, and I. Fried, "Single-neuron responses in humans during execution and observation of actions." Current biology (2010), 20(8), 750-756.